\documentclass[twocolumn,pra,superscriptaddress]{revtex4-2}
\usepackage{amsmath,amssymb,bm}
\usepackage{xcolor}
\usepackage{graphicx,subfigure}
\usepackage{BOONDOX-cal}          	
\usepackage{dsfont}					
\usepackage{easyReview}             
\usepackage{siunitx}				
\usepackage{physics}				
\usepackage{hyperref}				

\begin{document}
\title{Geometric phase assisted enhancement of non-inertial cavity-QED effects}
\author{Navdeep Arya}
\email{navdeeparya.me@gmail.com}
\affiliation{Department of Physical Sciences, Indian Institute of Science Education \& Research (IISER) Mohali, Sector 81 SAS Nagar, Manauli PO 140306 Punjab, India}

\author{Vikash Mittal}
\email{vikashmittal.iiser@gmail.com}
\affiliation{Department of Physical Sciences, Indian Institute of Science Education \& Research (IISER) Mohali, Sector 81 SAS Nagar, Manauli PO 140306 Punjab, India}

\author{Kinjalk Lochan}
\email{kinjalk@iisermohali.ac.in}
\affiliation{Department of Physical Sciences, Indian Institute of Science Education \& Research (IISER) Mohali, Sector 81 SAS Nagar, Manauli PO 140306 Punjab, India}

\author{Sandeep K.~Goyal}
\email{skgoyal@iisermohali.ac.in}
\affiliation{Department of Physical Sciences, Indian Institute of Science Education \& Research (IISER) Mohali, Sector 81 SAS Nagar, Manauli PO 140306 Punjab, India}

\begin{abstract}
	The state of a quantum system acquires a phase factor, called the geometric phase, when taken around a closed trajectory in the parameter space, which depends only on the geometry of the parameter space. Due to its sensitive nature, the geometric phase is instrumental in capturing weak effects such as the acceleration-induced non-inertial quantum field theoretic effects. In this paper, we study the geometric phase response of a circularly rotating detector inside an electromagnetic cavity. Using the cavity, the non-inertial contribution to the geometric phase can be isolated from or strengthened relative to the inertial contribution. We show that the accumulative nature of the geometric phase may facilitate the experimental observation of the resulting, otherwise feeble, non-inertial contribution to the modified field correlations inside the cavity. Specifically, we show that the atom acquires an experimentally detectable geometric phase at accelerations of the order of $\sim 10^{7} \si{m/s^2}$ which is experimentally feasible.
\end{abstract}

\maketitle

\section{Introduction}
The experimental detection of the non-inertial effects in quantum field theory, e.g., the Unruh effect~\cite{fulling1973,davies1975,unruh1976,crispino2008}, the Hawking radiation~\cite{HawkingG1974, Hawking1975}, has remained elusive till date because of the requirement of extreme conditions. For example, in the Unruh effect, to perceive a thermal effect of $1\si{K}$, typically an acceleration of the order of $10^{21}\si{m/s^2}$ is required~\cite{crispino2008}. Numerous attempts have been made to relax extreme conditions using sophisticated techniques and precise measurements, but only to a limited success~\cite{Rogers1988,Tajima1999,fuentes2010,scully2003,Lochan2021,Vanzella2001,Barshay1978,*Barshay1980,*Kharzeev2006,bell1983,*bell1987,*unruh1998}.

It has been further advocated that quantum features such as geometric phase (GP) may be of much usage in bringing down the scales involved. A quantum system taken around a closed path by varying the parameters $\vb{R}$ of its Hamiltonian $H(\vb{R})$ acquires a phase that is geometric in nature.
This phase is different from the standard dynamical phase of quantum systems in the sense that it depends only on the geometry of the parameter space~\cite{Berry1984}. In a quantum system interacting with an environment, GP depends not only on the unitary evolution, but also on transition and decoherence rates~\cite{Carollo2003}. Moreover, due to its accumulative and sensitive nature, the GP can be helpful in capturing weak effects such as the non-inertial effects in quantum field theory.

It has been proposed~\cite{Martinez2011} that using the GP the Unruh effect can be detected for accelerations as low as  $10^{17}\si{m/s^2}$, which, though a significant reduction in the acceleration typically required, is still extremely challenging to achieve. In~\cite{HuYu2014}, the GP acquired by a circularly rotating detector in free space was studied, but the detector's non-inertial response remains too feeble to be detectable at physically realizable accelerations. 

Interestingly, the usage of an electromagnetic cavity has been argued to further relax the acceleration requirement by a few orders~\cite{scully2003,lochan2020,Lochan2021}.
For example, by studying the spontaneous emission rate of an atom circularly rotating inside an electromagnetic cavity, it has been shown that the non-inertial effects can be detected at accelerations as low as $10^{12}\si{m/s^2}$~\cite{lochan2020}. 

In this paper, we study the GP acquired by a circularly rotating two-level atom inside an electromagnetic cavity. Since the GP is sensitive to transition rates and we already know that the transition rates become significantly modified inside the cavity~\cite{Purcell1995}, we expect the non-inertial component of the GP to be correspondingly modified. The accumulative nature of the GP~\cite{Berry1984,Tong2004} facilitates the detection of weak effects such as the non-inertial modifications to the field correlators, whereas using the electromagnetic cavity the atom's non-inertial response can be isolated from, or strengthened relative to, the inertial response~\cite{lochan2020}.

By studying the GP response of the atom inside an electromagnetic cavity, we show that the acceleration-induced modifications to the field correlators can be detected at much lower accelerations and with much relaxed parameters for the experimental setup. Specifically, we show that the atom acquires an experimentally detectable GP at accelerations of the order of $10^{7} \si{m/s^2}$ which is experimentally feasible. Therefore, an efficient GP measurement inside a cavity is supposed to manifest the non-inertial quantum field theoretic effects more robustly as compared to any other setup proposed so far.

The paper is organized as follows. In Sec.~\ref{Sec:GP in open quantum system} we introduce the background relevant to understanding our results. We begin by discussing the reduced dynamics of the two-level system interacting with an external environment and introduce the notion of GP for mixed states. In Sec.~\ref{Sec:GP response of the rotating detector}, we present our results on the GP response of the circularly rotating detector. We conclude and discuss our results in Sec.~\ref{Sec:Conclusion and Discussion}.   

\section{GP in open quantum system}\label{Sec:GP in open quantum system}
In this section we introduce the concept of GP for an open quantum system. For that, consider a two-level atom interacting with an electromagnetic field. The atomic ground and excited states are denoted, respectively, by $\ket{g}$ and $\ket{e}$. The proper frequency gap between the two atomic levels is $\Omega_0$ and the atom carries an electric dipole moment four-vector $
\hat{d}'^{\mu}$. Throughout this paper, primed quantities refer to the atom's co-moving frame. In the interaction picture, the dipole moment operator $\hat{\vb{d}}'(\tau)$ is given in terms of its matrix elements (as they will appear in later expressions) by
\begin{equation}\label{eq:dipole_operator}
	\hat{\vb{d}}'(\tau) = \vb{d}' \sigma_{-} \exp(-i \Omega_0 \tau) + \vb{d}'^* \sigma_{+} \exp(i \Omega_0 \tau), 
\end{equation} 
where $\vb{d}' \equiv \bra{g} \hat{\vb{d}}'(\tau = 0) \ket{e}$ and $\sigma_{+} = \sigma^{\dagger}_{-} = \ket{e} \bra{g}$ is the step-up operator for the atomic states. For simplicity, we assume that $\vb{d}' = (0, d', 0)$.  The electromagnetic field is assumed to be in the Minkowski vacuum state $\ket{0}$. The interaction Hamiltonian between the atom and the electromagnetic field is given by $H_I = - d^{\mu}E_{\mu}$~\cite{Anandan2000}, where $E_{\mu} \equiv F_{\mu \nu} u^{\nu}$, $F_{\mu\nu}$ is the electromagnetic field strength tensor and $u^{\nu}$ is the four-velocity of the atom. The interaction Hamiltonian takes the form $H_I = - \vb{d'}\cdot \vb{E'}$ in the rest frame of the atom, where $\vb{E}'$ is the electric field 3-vector as seen by the atom. 
The electric field operator inside a quantization volume $V$ is given by~\cite{Gerry2004}
\begin{equation}
	\vb{E}[x(\tau)] = i \sum_{\vb{k},\lambda} \mathcal{E}_k \epsilon_{\vb{k},\lambda} \left(a_{\vb{k},\lambda} e^{-i\left(\omega_k t(\tau) - \vb{k}.\vb{x}(\tau)\right)}- \text{h.c.}\right),
\end{equation}
where $\mathcal{E}_k \equiv \sqrt{\hbar \omega_k/(2\epsilon_0 V)}$, $\epsilon_{\vb{k},\lambda}$ with $\lambda = 1,2$ are the two orthogonal polarization vectors, and h.c. denotes Hermitian conjugate.

In the rest frame of the atom, the evolution of the atom-field composite system is governed by \cite{Breuer2007} 
\begin{equation}\label{total_time_evolution}
	\derivative{\rho_{\text{AF}}(\tau)}{\tau} = - \frac{i}{\hbar} \comm{H}{\rho_{\text{AF}}(\tau)},
\end{equation} 
where $\rho_{\text{AF}}$ is the density operator of the composite system and $\tau$ is the proper time of the atom. From (\ref{total_time_evolution}), we can obtain the Lindblad evolution of the reduced density operator for the atom $\rho(\tau) \equiv \Tr_{F}{\left(\rho_{\text{AF}}\right)}$, which is given by \cite{Lindblad1976}
\begin{equation}\label{lindblad_eqn}
	\derivative{\rho(\tau)}{\tau} = -\frac{i}{\hbar} \comm{H_{\text{eff}}}{\rho(\tau)} + \mathcal{L}[\rho(\tau)],
\end{equation} 
where 
\begin{equation} \label{eq:lindbladoperator}
	\mathcal{L}[\rho] = \frac{1}{2} \sum_{i,j = 1}^{3} a_{ij} \left(2 \sigma_j \rho \sigma_i - \sigma_i \sigma_j \rho - \rho \sigma_i \sigma_j \right),
\end{equation} 
captures the dissipation and decoherence of the atom induced by its interaction with the electromagnetic field. The $H_{\text{eff}}$ represents the Hamiltonian of the two-level atom with the renormalized atomic level spacing, $ \Omega $, which consists of the Lamb shift~\cite{Breuer2007}. The $\sigma_i$'s are the standard Pauli matrices~\cite{Sakurai2017} and the coefficients $a_{ij}$ are given by \cite{Benatti2005} 
\begin{equation} \label{eq:Kosskowskimatrix}
	a_{ij} = A \delta_{ij} - i B \epsilon_{ijk} \delta_{k3} - A \delta_{i3} \delta_{j3},
\end{equation} 
with $A$ and $B$ defined as 
\begin{align}\label{eq:AB}
	A &= \frac{1}{4} \left[\Gamma_{\downarrow} + \Gamma_{\uparrow}\right], \;\;\; B = \frac{1}{4} \left[\Gamma_{\downarrow} - \Gamma_{\uparrow}\right],
\end{align} 
whereas,
\begin{equation} \label{FTCF}
	\begin{split}
		\Gamma_{\downarrow \uparrow} &= \abs{\bra{\psi_f} \hat{\vb{d}}'(0) \ket{\psi_i}}^2 \int_{-\infty}^{\infty} \dd{\tau_-} e^{\pm i \Omega_0 \tau_-} G'^{+}(x(\tau_-)),
	\end{split}
\end{equation} 
with `$+$' and `$-$' corresponding to $\Gamma_{\downarrow}$ and $\Gamma_{\uparrow}$, i.e., the emission and the absorption rates, respectively. Here, $\tau_- \equiv \tau_2 - \tau_1$; $\ket{\psi_i}, \ket{\psi_f} \in \{\ket{g}, \ket{e} \}$; and $G'^+(x(\tau_-)) \equiv \bra{0}E'^{y}(\tau_1)E'^{y}(\tau_2)\ket{0}$ is the two-point vacuum Wightman function. 
By taking the initial state of the atom to be
\begin{equation}\label{atomic_initial_state}
	\ket{\psi(0)} = \cos(\theta/2) \ket{e} + \sin(\theta/2) \ket{g},
\end{equation} 
and solving Eq.~\eqref{lindblad_eqn}, we get the reduced density operator to be
\begin{widetext} 
	\begin{equation} \label{eq:reduceddensity}
		\rho(\tau) = \begin{pmatrix}
			e^{-4A\tau}\cos^2(\theta/2) + \frac{B-A}{2A}\left(e^{-4A\tau}-1\right) & \frac{1}{2} e^{-2A\tau-i\Omega \tau}\sin\theta \\ \noalign{\vskip7pt}
			\frac{1}{2} e^{-2A\tau+i\Omega \tau}\sin\theta & 1 - e^{-4A\tau}\cos^2(\theta/2) - \frac{B-A}{2A}\left(e^{-4A\tau}-1\right)\\
		\end{pmatrix},
	\end{equation}
	in which the effect of the environment is contained in $A$ and $B$.  The GP for an $N$-level quantum system in a mixed state and evolving non-unitarily is given by \cite{Tong2004} 
	\begin{align}
		\gamma_g
		=\arg \left( \sum_{k=1}^{N} \sqrt{p_k(0)p_k(T)} \ip*{\phi_k(0)}{\phi_k(T)} e^{-\int_0^{T} \ip*{\phi_k(\tau)}{\dot{\phi}_k(\tau)} d\tau } \right) \label{eq:tong},
	\end{align}
\end{widetext}
where $p_k(\tau)$ and $\ket{\phi_k(\tau)}$ are instantaneous eigenvalues and eigenvectors of the system's density operator $\rho(\tau)$. The eigenvalues of $\rho(\tau)$ are
\begin{equation} \label{eq:eigenvalue}
	p_{\pm}(\tau) = \dfrac{1}{2} \left( 1 \pm \lambda\right),
\end{equation}
where $\lambda = \sqrt{r_3^2 + e^{-4A}\sin^2\theta}$ and $r_3 = e^{-4A\tau}\cos\theta + \tfrac{B}{A}(e^{-4A\tau} -1)$. Since $p_-(0) = 0$, the only eigenvector that contributes to $\gamma_g$  is the one corresponding to $p_+$ which reads
\begin{equation} \label{eq:eigenvector}
	\ket{\phi_+(\tau)} = \sin (\theta_{\tau}/2) \ket{+} + e^{i \Omega_0 \tau} \cos (\theta_{\tau}/2) \ket{-},
\end{equation}
with
\begin{equation}
	\tan (\theta_{\tau}/2) = \sqrt{\dfrac{\lambda + r_3}{\lambda - r_3}}.
\end{equation}
By substituting Eq.~\eqref{eq:eigenvalue} and Eq.~\eqref{eq:eigenvector} in the expression of the GP in Eq.~\eqref{eq:tong}, we get \cite{HuYu2012}
\begin{align} 
	\gamma_g &= - \dfrac{\Omega}{2} \nonumber \\ 
	&\times \int_{0}^{T} \dd{\tau} \left(1 - \dfrac{\mathcal{R} - \mathcal{R} e^{4 A \tau} + \cos \theta}{\sqrt{e^{4 A \tau} \sin^2 \theta + (\mathcal{R} - \mathcal{R} e^{4 A \tau} + \cos \theta)^2}}\right) \label{eq:geophase},
\end{align}
where $\mathcal{R} \equiv B/A$. The above expression is valid for all time $T$. Now, if we take $T = 2 \pi n/\Omega_0$, where $n$ is the number of quasi-cycles, and if $A/\Omega_0 \ll 1$, we can further simplify the expression for $\gamma_g$ and obtain~\cite{HuYu2012}
\begin{equation} \label{eq:HuYuGamma}
	\gamma_g = -\pi n (1 - \cos \theta) - \dfrac{2 \pi^2 n^2}{\Omega_0} \left( 2 B + A \cos \theta \right) \sin^2 \theta,
\end{equation} 
where $n$ is restricted by the demand $\pi n A/\Omega_0 \ll 1$, because to obtain \eqref{eq:HuYuGamma} from \eqref{eq:geophase}, we need $4A\phi/\Omega_0 \ll 1$ and for $n$ cycles $\phi = 2\pi n$. The GP has two contributions: the first term in Eq.~\eqref{eq:HuYuGamma} is due to the unitary evolution of the atom and the second term is coming from the non-unitary evolution of the atom which results from the atom's interaction with the environment. We will shortly see that in case of non-inertial motion of the atom, the non-unitary contribution can be further separated into an inertial and a non-inertial part. Further, using the cavity's resonance structure, the non-inertial contribution to the GP can be significantly enhanced.

\section{GP response of the circularly rotating detector}\label{Sec:GP response of the rotating detector}
In this section, we study the GP acquired by an atom moving on a circular trajectory of radius $R$ and angular frequency $\omega$, inside an electromagnetic cavity. The rotating $(\tau, x', y', z')$ and the inertial coordinates ($t, x, y, z$) are related as 
\begin{align}
	x' & = x - x_0 - R \cos\omega t, \nonumber \\ \noalign{\vskip5pt}
	y' & = y,\nonumber  \\ \noalign{\vskip5pt}
	z' & = z - z_0 - R \sin\omega t, \nonumber  \\ \noalign{\vskip5pt}
	\tau &= \left(1-\omega^2 R^2/c^2\right)^{1/2} t, \label{eq:coordinates}
\end{align} 
where $(x_0, 0, z_0)$ is the center of the circular trajectory. 
We have seen in Eq.~\eqref{eq:HuYuGamma} that the GP depends directly on the quantities $A$ and $B$ which, in turn, depend on the atomic transition rates. The non-inertial motion of the atom only affects the transition rates and leaves the form of Lindblad master equation [Eq.~\eqref{lindblad_eqn}] unchanged. Therefore, the expression of GP in Eq.~\eqref{eq:geophase} can be used in the case of a rotating atom with the modified transition rates. It is natural to start, therefore, by computing the transition rates and then obtain the GP. 

We study the GP in two different regimes distinguished by the relative magnitudes of the rotational frequency of the atom $(\omega)$ and the atomic frequency gap $(\Omega_0)$. Specifically, we study the $\omega >> \bar{\Omega}_0$ and $\omega << \bar{\Omega}_0$ regimes, where $\bar{\Omega}_0 \equiv \Omega_0 \sqrt{1-\omega^2R^2/c^2}$.
As we shall see, in the $\omega >> \bar{\Omega}_0$ regime, the non-inertial contribution in the spontaneous decay rate dominates over the inertial one when the normal frequency of the cavity is tuned at $\omega+\bar{\Omega}_0$. In the $\omega << \bar{\Omega}_0$ regime, on the other hand, the inertial contribution to the GP overshadows the non-inertial component. However, with a suitable choice of parameters, the non-inertial contribution can be made comparable to the inertial component. Therefore, in both regimes, we can analyze the non-inertial contribution to the GP effectively.

\subsection{Transition rates in the atom's frame}
If the atom-field composite system is initially in the state $\ket{e,0}$, then the total probability of spontaneous emission, in the co-moving frame, using Born's rule is given, to the leading order in atomic dipole moment, by 
\begin{equation}
	\begin{split}\label{eq:spon_decay_prob}
		P_{\downarrow} &= \sum_{\Psi}\abs{\bra{g,\Psi}\left[\frac{-i}{ \hbar}\int_{-\infty}^{\infty}\dd{\tau}H_I(\tau)\right]\ket{e,0}}^2 \\\noalign{\vskip5pt}
		&\equiv \frac{1}{\hbar^2} \int \dd{\tau_1} \int \dd{\tau_2}  \bra{e} \hat{d}'^{\mu}(\tau_2) \ket{g} \bra{g} \hat{d}'^{\nu}(\tau_1) \ket{e} G'^{+}_{\mu \nu}, \\
		& = \frac{\abs{\vb{d}'}^2}{\hbar^2} \int \dd{\tau_1} \int \dd{\tau_2}  e^{i\Omega_0 (\tau_2 - \tau_1)} G'^{+}_{22},
	\end{split}
\end{equation} 
where primed quantities are the quantities as seen from the co-moving frame, $G'^{+}_{\mu \nu} \equiv \bra{0} E'_{\mu}(\tau_2) E'_{\nu}(\tau_1) \ket{0}$ is the two-point vacuum Wightman tensor, and $\ket{\Psi}$ are different possible final states of the field. $G'^{+}_{22}$ can be obtained from its inertial counterpart, using  Eq.~\eqref{eq:coordinates} as
\begin{equation}
	G'^{+}_{22} = \sum_{\mu \nu} \pdv{x^{\mu}}{x'^{2}} \pdv{x^{\nu}}{x'^{2}} G^{+}_{\mu \nu} =  G^{+}_{22},
\end{equation}
where
\begin{equation}\label{eq:atomic_four_vec}
	\begin{split}
		&x^{\mu}(\tau) \\
		&\;= (t(\tau), x(\tau), y(\tau), z(\tau))\\
		&\;= \left(c\gamma \tau, x_0 + R \cos(\omega \gamma \tau), 0, z_0 + R \sin(\omega \gamma \tau)\right),
	\end{split}
\end{equation}
is the atomic position four-vector in the lab frame, with $\gamma \equiv (1-\zeta(\omega))^{-1/2}$, and $\zeta(\omega) \equiv \omega^2 R^2/c^2$.
The electric field perceived by the atom and as reported by the inertial observer is given by $E_{\mu}(x^{\lambda}) = F_{\mu \nu}(x^{\lambda}) u^{\nu}$, where $u^{\nu} \equiv \dv*{x^{\nu}(\tau)}{\tau}$ is the atomic four-velocity in the lab frame. Therefore, using Eq.~\eqref{eq:atomic_four_vec} we have
\begin{equation}\label{E_2}
	\begin{split}
		E_2 =\gamma \left[ E_y - \omega R \left\{ B_z \sin(\omega \gamma \tau) - B_x \cos(\omega \gamma \tau) \right\} \right].
	\end{split}
\end{equation}
Further, using Eq.~\eqref{E_2} we can obtain $G^{+}_{22}$ and consequently, with Eq.~\eqref{eq:spon_decay_prob}, and writing $\dd \tau = \dd t/\gamma$, we obtain
\begin{multline}\label{decay_prob_1}
	P_{\downarrow} = \frac{\abs{\vb{d}}^2}{ \gamma^2 \hbar^2} \int \dd{t_1} \int \dd{t_2}~ e^{i\bar{\Omega}_0(t_2-t_1)}  \bra{0}\gamma [E_{y}(x^{\mu}_1) \\ - R\omega \sin(\omega t_1)B_{z}(x^{\mu}_1)- R\omega \cos(\omega t_1) B_{x}(x^{\mu}_1)] \times \gamma [E_{y}(x^{\mu}_2) \\ - R\omega \sin(\omega t_2)B_{z}(x^{\mu}_2) - R\omega \cos(\omega t_2) B_{x}(x^{\mu}_2)] \ket{0},
\end{multline}
where $\vb{E} = \left(E_x, E_y, E_z \right)$ and $\vb{B} = \left(B_x, B_y, B_z \right)$ are the electric and magnetic field 3-vectors, respectively, as seen by static observers in the lab frame and $x^{\mu}_1, x^{\mu}_2$ are two spacetime points.
Similarly, we can compute the absorption probability $(P_{\uparrow})$ which is obtained by changing $\bar{\Omega}_0$ in Eq.~\eqref{decay_prob_1} to $-\bar{\Omega}_0$.

Effecting a change of time variables, $t_{+} \equiv \left(t_{1} + t_{2}\right)/2$, and $t_{-} \equiv t_{2} - t_{1}$, and evaluating the correlators~\cite{lochan2020}, we get the transition rates in the lab frame for small $\zeta(\omega)$, and to the first order in $\zeta(\omega)$, as
\begin{widetext}
	\begin{equation}\label{transition_rates}
		\begin{split}
			\Gamma^{\text{lab}}_{\downarrow \uparrow} = &\eta \int \dd{k}\rho(k)  \omega_k \Bigg[\delta(\pm \bar{\Omega}_0 - \omega_k) +\frac{R^2 \omega^2}{2c^2}  \frac{ 1}{2} \left[\delta(\pm \bar{\Omega}_0  - \omega_k + \omega) + \delta(\pm \bar{\Omega}_0  - \omega_k - \omega)\right] \\  &\qquad\qquad- \frac{\omega_k^2 R^2}{c^2} \frac{2}{5} \left\{ \delta(\pm \bar{\Omega}_0 - \omega_k)  - \frac{1}{2}  \left(\delta(\pm \bar{\Omega}_0 + \omega -\omega_k) + \delta(\pm \bar{\Omega}_0 - \omega -\omega_k)\right) \right\} \Bigg],
		\end{split}
	\end{equation}
\end{widetext} 
where $\eta \equiv \abs{\vb{d}}^2/(3 \pi \hbar \epsilon_0 V)$; `$+$' and `$-$' correspond, respectively, to the spontaneous decay rate $(\Gamma_{\downarrow})$ and the excitation rate $(\Gamma_{\uparrow})$ with $\rho(k)$ being the density of the field states. The transition rates in the co-moving frame can be obtained as $ \Gamma_{\downarrow \uparrow} = \gamma \Gamma^{\text{lab}}_{\downarrow \uparrow}$. Inside a cavity, the density of states can be taken to be of Lorentzian form~\cite{Agarwal2012}
\begin{align}
	\rho(\omega_k) \sim \frac{\left(\omega_c/Q\right)}{\left(\omega_c/Q\right)^2 + \left(\omega_k - \omega_c\right)^2},
\end{align} 
where $\omega_c$ is the normal frequency and $Q$ is the quality factor of the cavity. 
The expression for the transition rates takes a simpler form in the two regimes, namely $\omega \gg \bar{\Omega}_0$ and $\omega \ll \bar{\Omega}_0$. Now, we shall explore the transition rates and the GP response of the atom in these two regimes.

\begin{figure*}
	\centering
	\subfigure[]{
		\includegraphics[height=4.3cm]{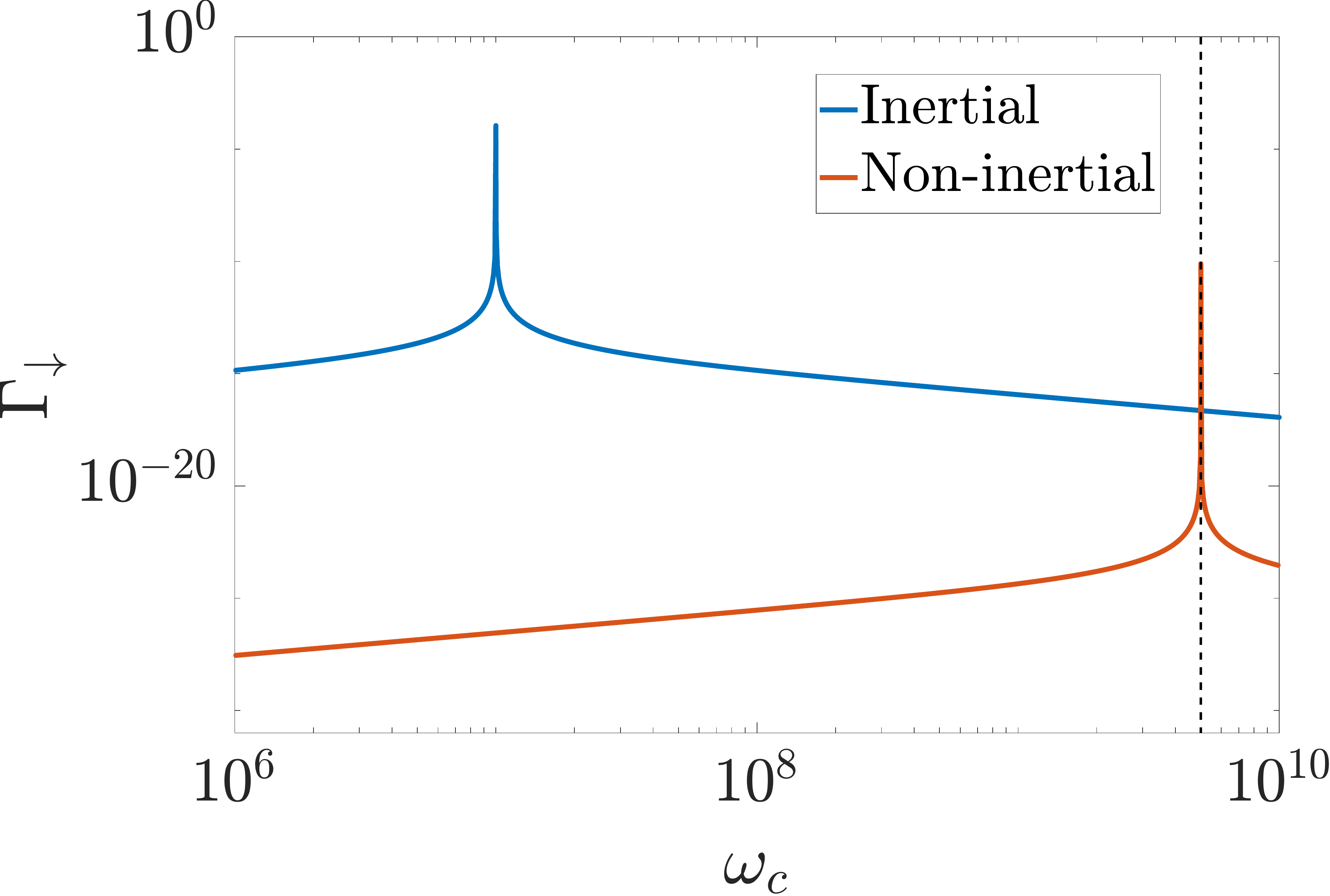}
		\label{fig:decayrate1}}
	\subfigure[]{
		\includegraphics[width=6.5cm]{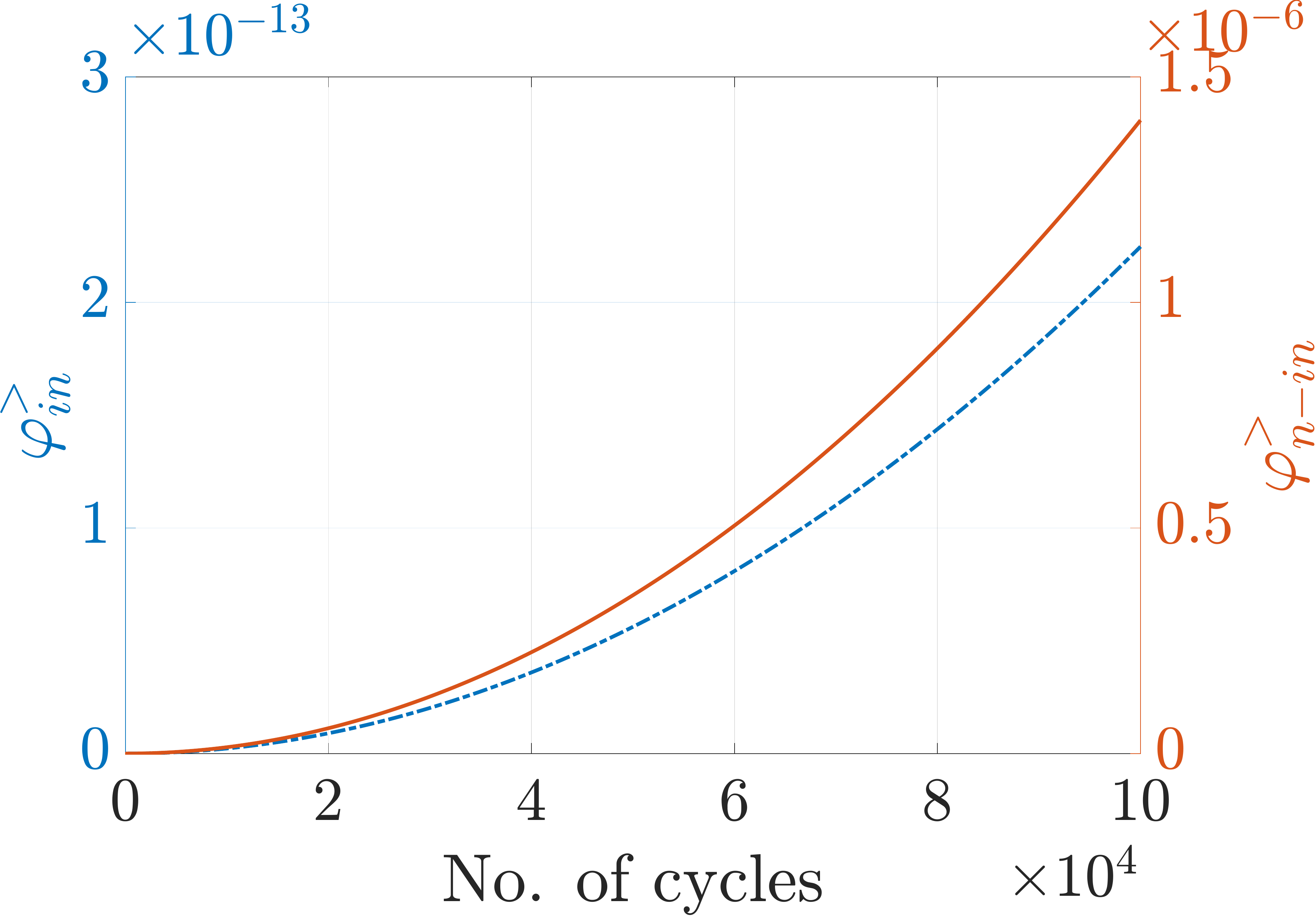}
		\label{fig:gp1}}
	\subfigure[]{
		\includegraphics[height=4.3cm]{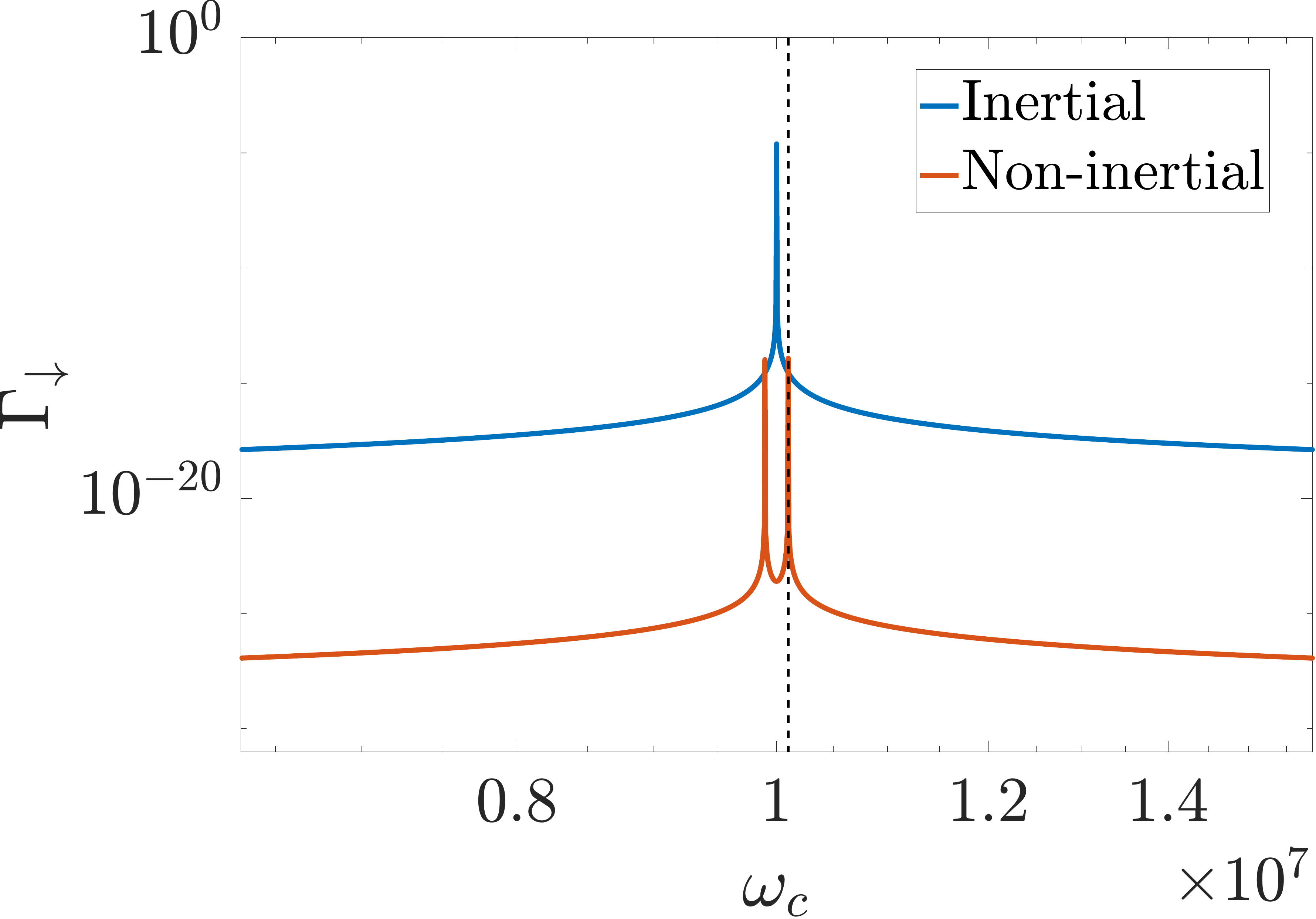}
		\label{fig:decayrate2}} 
	\subfigure[]{
		\includegraphics[width=6.5cm]{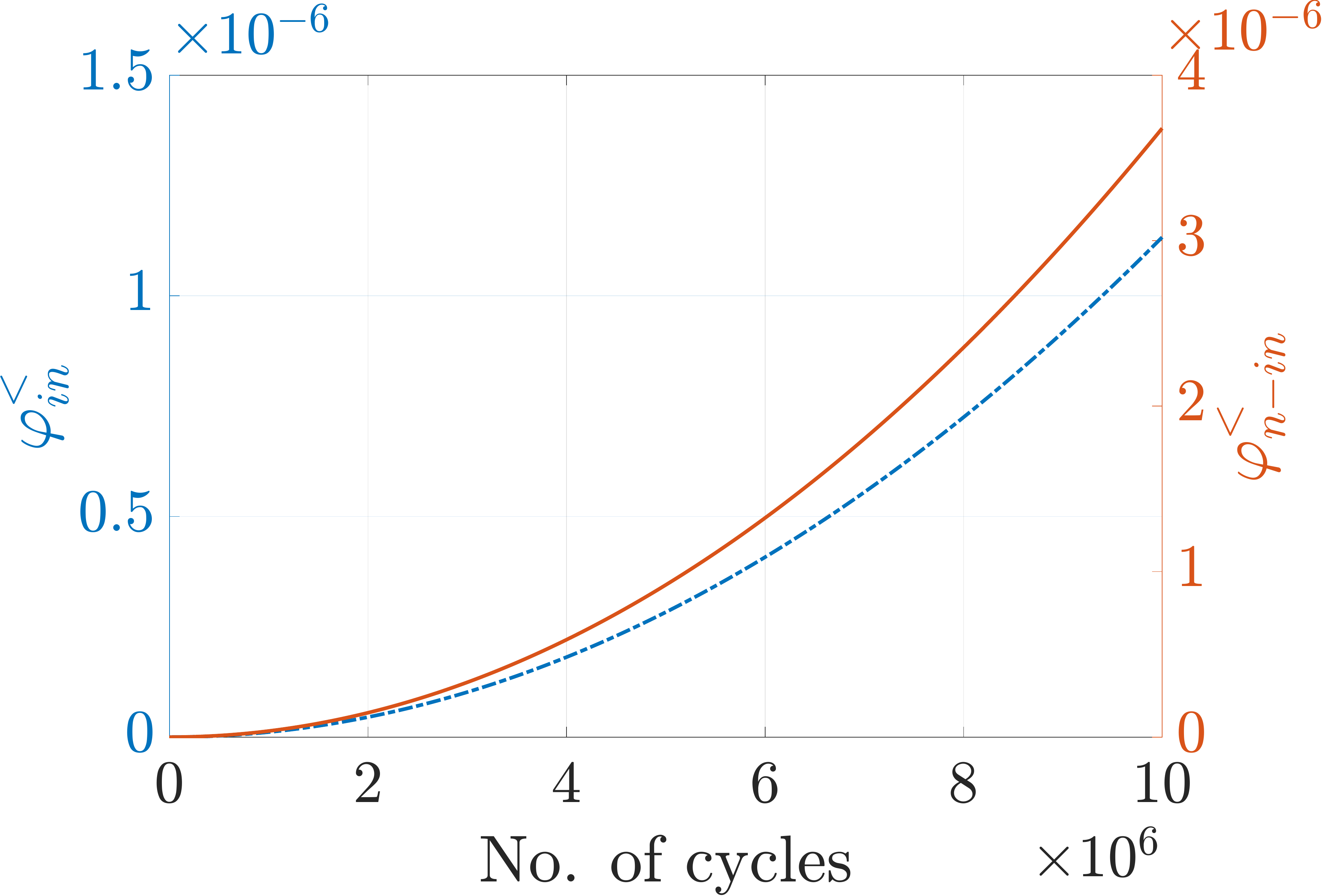}
		\label{fig:gp2}}
	\caption{(Color online) The plot for $\Gamma_{\downarrow}$ vs. $\omega_c$ and the non-unitary GP versus the number of quasi-cycles ($n$) in the two regimes discussed in the paper. We plot the inertial ($\varphi_{in}$) and the non-inertial contributions ($\varphi_{n-in}$) to the non-unitary GP here. \subref{fig:decayrate1},\subref{fig:gp1} $\omega >> \bar{\Omega}_0$ with $\omega = 5$ GHz, $\Omega_0 = $ 10 MHz, $V = 10^{-7} \si{m^3}$, and $R = 10^{-6} \si{m}$ which correspond to an average acceleration $a = \omega^2 R \sim 2.5 \times 10^{13} \si{m/s^2}$. For this set of parameters, $\pi n A/\Omega_0 \sim 10^{-16} n$. \subref{fig:decayrate2},\subref{fig:gp2} $\omega << \bar{\Omega}_0$ with $\omega = $ 0.1 MHz, $\Omega_0 = $ 10 MHz, $V = 10^{-3} \si{m^3}$, and $R = 10^{-3} \si{m}$ which correspond to an average acceleration $a = \omega^2 R \sim 10^{7} \si{m/s^2}$.The plots are for $\theta = \pi/2$ in Eq.~\eqref{atomic_initial_state}. The vertical black dashed lines in Figs.~\subref{fig:decayrate1},~\subref{fig:decayrate2} mark the normal frequencies at which the cavity is tuned to obtain the GP versus $n$ plots in Figs.~\subref{fig:gp1},\subref{fig:gp2}.}
	\label{fig:case2}
\end{figure*}

\subsubsection*{Case 1: $\omega \gg \bar{\Omega}_0$}
Using Eq.~\eqref{transition_rates} and $ \Gamma_{\downarrow \uparrow} = \gamma \Gamma^{\text{lab}}_{\downarrow \uparrow}$, the spontaneous emission rate in the $\omega \gg \bar{\Omega}_0$ regime can be obtained as 
\begin{multline}\label{eq:Gamma_down}
	\Gamma_{\downarrow} =  \eta \gamma \Big[\bar{\Omega}_0 \left(1 - \frac{2}{5}\zeta(\bar{\Omega}_0)\right) \rho(\bar{\Omega}_0) + \frac{9}{20} \zeta(\omega) \omega_+ \rho(\omega_+)\Big].
\end{multline}
By expanding the first term in \eqref{eq:Gamma_down} to the first order in $\zeta(\omega)$ and using $\gamma = 1 + \zeta(\omega)/2 + \order{\zeta(\omega)^2}$, we can write the spontaneous emission rate to the first order in $\zeta(\omega)$ as
\begin{equation}
	\Gamma_{\downarrow} = \eta \rho(\Omega_0) \Omega_0  + \Gamma_{\downarrow}^{\text{ni}},
\end{equation}
where the first term is the inertial contribution, obtained by taking the $\omega \rightarrow 0$ limit, and the second term is the purely non-inertial contribution coming through the rotation.
The purely non-inertial contribution to the spontaneous decay rate can be obtained as follows:
\begin{equation}\label{purely_ni_decayrate}
	\Gamma_{\downarrow}^{\text{ni}} = \Gamma_{\downarrow} - \Gamma_{\downarrow}(\omega \to 0),
\end{equation} 
which, to the first order in $\zeta(\omega)$, turns out to be 
\begin{equation}
	\begin{split}
		\Gamma_{\downarrow}^{\text{ni}} &= \frac{\eta \zeta(\omega)\Omega_0}{2}  \Bigg[- \Omega_0 \rho'(\Omega_0) + \frac{9}{10}  \frac{\omega_+}{\Omega_0} \rho(\omega_+) \Bigg]
	\end{split}
\end{equation}
where $\rho'(\omega) = \partial \rho / \partial \omega$.

Again, using Eq.~\eqref{transition_rates} and $ \Gamma_{\downarrow \uparrow} = \gamma \Gamma^{\text{lab}}_{\downarrow \uparrow}$, the absorption rate in the $\omega \gg \bar{\Omega}_0$ regime turns out to be
\begin{equation}
	\Gamma_{\uparrow} = \frac{9}{20} \eta \gamma \zeta(\omega) \rho(\omega_-) \omega_-,
\end{equation} 
where $\omega_{\pm} \equiv \omega \pm \bar{\Omega}_0$. Note that, as expected, the absorption rate has only non-inertial contribution. Furthermore, using $\gamma = 1 + \zeta(\omega)/2 + \order{\zeta(\omega)^2}$, the absorption rate to the first order in $\zeta(\omega)$ is given by
\begin{equation}\label{eq:Gamma_up}
	\Gamma_{\uparrow} = \frac{9}{20} \eta \zeta(\omega) \rho(\omega_-) \omega_-.
\end{equation}
From Eq.~\eqref{eq:AB}, we know that $A$ is the sum of the spontaneous emission and absorption rates and, therefore, up to the first order in $\zeta(\omega)$, is given by
\begin{widetext}
	\begin{equation}\label{eq:Acase1}
		A = \frac{\eta}{4}  \Bigg[ \rho(\Omega_0) \Omega_0 - \frac{\zeta(\omega)}{2} \Omega_0^2 \rho'(\Omega_0) + \frac{9}{20} \zeta(\omega) \left\{ \omega_+ \rho(\omega_+) +  \rho(\omega_-) \omega_- \right\}\Bigg].
	\end{equation}
	Similarly, $B$ is obtained by replacing the term in curly brackets in Eq.~\eqref{eq:Acase1} by $\omega_+ \rho(\omega_+) -  \rho(\omega_-) \omega_-$. Using \eqref{eq:HuYuGamma}, we obtain the non-unitary GP acquired by the atom in $n$ number of quasi-cycles to be
	\begin{multline} \label{eq:Gamma_g_high}
		\varphi^{>} = - \dfrac{2 \pi^2 n^2}{\Omega_0} \frac{\eta}{4} \Bigg[ \left(\rho(\Omega_0) \Omega_0 - \frac{\zeta(\omega)}{2} \Omega_0^2 \rho'(\Omega_0)\right) (2 + \cos\theta) + \frac{9}{20} \zeta(\omega) \Bigg\{ \omega_+ \rho(\omega_+)  (2 + \cos\theta) - \omega_- \rho(\omega_-) (2 - \cos\theta) \Bigg\} \Bigg] \sin^2 \theta,
	\end{multline}
\end{widetext}
where a quasi-cycle consists of a time period of $T = 2\pi/\Omega_0$.
Note that we are using the symbol $\varphi^{>}$ to denote the non-unitary GP in the $\omega \gg \bar{\Omega}_0$ regime. Next, we discuss some numerical estimates of GP for realistic settings~\cite{Ahn2018,Leduc2010,lochan2020}.

For $\Omega_0 = 10~\si{MHz}$, $Q = 10^7$, and $\omega = 5~\si{GHz}$, when $\omega_c = \omega_+$, that is, when the cavity is tuned to $\bar{\Omega}_0 + \omega$, we have $\rho(\Omega_0) \sim 10^{-14}, \; \rho'(\Omega_0) \sim 10^{-7}, \; \rho(\omega_+) \sim 10^{-2}$, and $\rho(\omega_-) \sim 10^{-13}$, thus making the non-inertial contribution to $A$ and $B$ highly dominant over the inertial contribution. The parameters mentioned above lead to an average acceleration $a = \omega^2 R \sim 10^{13}~\si{m/s^2}$.  Fig.~\ref{fig:decayrate1} illustrates the variation of the inertial and the non-inertial contributions to the spontaneous decay rate as the cavity is tuned to different frequencies. When we use $A$ and $B$ obtained by tuning the cavity to $\omega_+$, the non-unitary GP [see Eq.~\eqref{eq:Gamma_g_high}] acquired by the atom is predominantly non-inertial [see Fig.~\ref{fig:gp1}]. In Fig.~\ref{fig:gp1} we plot the inertial and the non-inertial contributions to the GP acquired by the atom as a function of the number of quasi-cycles for $V = 10^{-7} \si{m^3}$ and $R = 10^{-6} \si{m}$. With these values for the parameters, $\pi A n/\Omega_0 \sim 10^{-16}n$ which decides the allowed number of quasi-cycles consistent with the approximation made to obtain Eq.~\eqref{eq:HuYuGamma}. We note from Fig.~\ref{fig:gp1} that the system acquires an experimentally observable~\cite{Wang2018} non-inertial GP $\sim 10^{-6} ~\si{radian}$ (1 second radian) in roughly $10^5$ quasi-cycles, whereas the inertial contribution to the GP is $\sim 10^{-13} ~\si{radian}$ in the same number of quasi-cycles. 

\subsubsection*{Case 2: $\omega \ll \bar{\Omega}_0$}
If $\omega \ll \bar{\Omega}_0$, the spontaneous emission rate can be approximated, to the first order in $\zeta(\omega)$, to be
\begin{widetext}
	\begin{multline} \label{eq:sdecayratecase2}
		\Gamma_{\downarrow} = \eta  \Bigg[ \rho(\Omega_0)  \Omega_0 - \frac{\zeta(\omega)}{2} \Omega_0^2  \rho'(\Omega_0) + \frac{\zeta(\omega)}{4} \left( \rho(\bar{\Omega}^+_0)  \bar{\Omega}^+_0  + \rho(\bar{\Omega}^-_0)  \bar{\Omega}^-_0 \right) \\ - \frac{2}{5} \left(1+\frac{\zeta(\omega)}{2}\right) \left\{ \zeta(\bar{\Omega}_0) \rho(\bar{\Omega}_0)  \bar{\Omega}_0  - \frac{1}{2}  \left( \zeta(\bar{\Omega}^+_0)\rho(\bar{\Omega}^+_0)  \bar{\Omega}^+_0 +  \zeta(\bar{\Omega}^-_0)\rho(\bar{\Omega}^-_0) \bar{\Omega}^-_0 \right) \right\} \Bigg],
	\end{multline}
\end{widetext}
where $\bar{\Omega}^{\pm}_0 = \bar{\Omega}_0 \pm \omega >0$.
The absorption rate to the first order in $\zeta(\omega)$ vanishes. Therefore, we have 
\begin{equation}
	A = B = \Gamma_{\downarrow}/4,
\end{equation} which leads to a non-unitary GP given by
\begin{equation} \label{eq:Gamma_g_low}
	\varphi^{<} = - \dfrac{\pi^2 n^2}{2\Omega_0} \Gamma_{\downarrow} \left( 2 +  \cos \theta \right) \sin^2 \theta,
\end{equation}
where we have used the symbol $\varphi^{<}$ to denote the non-unitary GP in the $\omega \ll \bar{\Omega}_0$ regime.

The GP acquired by the rotating atom can be measured using atom interferometry~\cite{RMP2009_Pritchard}. Atom interferometry employs the wave nature of atoms to detect the relative phase between any two atoms. To measure the non-inertial GP acquired by the rotating atom, an inertial reference atom can be used. Thus, the inertial GP cancels out between the two interferometer arms and the non-inertial GP leads to a shift in the interference fringes. It is clear that in order to ascribe the fringe shift to the non-inertial GP, the cancellation of the inertial GP should be better than the magnitude of the non-inertial GP. If the inertial and non-inertial phase contributions are comparable, this condition can be easily achieved. Next, we show that it is, in fact, possible, with a suitable choice of parameters, to make the two contributions comparable even in the $\omega \ll \bar{\Omega}_0$ regime. Note that the non-inertial GP response has terms like $\zeta(\bar{\Omega}^+_0)\rho(\bar{\Omega}^+_0)  \bar{\Omega}^+_0$ and $\zeta(\bar{\Omega}^-_0)\rho(\bar{\Omega}^-_0) \bar{\Omega}^-_0$ that dominate over the inertial as well as other non-inertial terms when the cavity is tuned to $\bar{\Omega}^+_0$ and $\bar{\Omega}^-_0$, respectively, for a suitable choice of $R$.
With the parameter set $\omega \sim 10^5 \si{Hz}$, $V = 10^{-3}\si{m^3}$, $R \sim 10^{-3} \si{m}$, if we tune the cavity to $\bar{\Omega}^+_0$, we have $\rho(\Omega_0) \sim 10^{-10}, \; \rho'(\Omega_0) \sim 10^{-5}, \rho(\bar{\Omega}^+_0) \sim 1,~\text{and}~ \rho(\bar{\Omega}^-_0) \sim 10^{-11}$ and the inertial and the non-inertial contributions to the transition rate become comparable [see Fig.~\ref{fig:decayrate2}]. Although with this parameter set, the resulting non-unitary GP per cycle, to which the inertial and non-inertial contributions are comparable, is extremely small, due to the accumulative nature of the GP it can be enhanced by allowing the atom to evolve for a higher number of quasi-cycles. With the parameter set considered here, the average acceleration turns out to be $ a = \omega^2 R \sim 10^7~\si{m/s^2} $. Fig.~\ref{fig:gp2} gives the plot of the inertial and the non-inertial contributions to the GP as a function of the number of quasi-cycles. With the parameters taken in Fig.~\ref{fig:gp2}, the allowed number of quasi-cycles consistent with Eq.~\eqref{eq:HuYuGamma} is determined by $\pi A n/\Omega_0 \ll 1$, that is, $10^{-21} n \ll 1$. 

\section{Discussion and Conclusion}\label{Sec:Conclusion and Discussion}
It is essential to note that although a circularly rotating detector does not perceive the Minkowski vacuum to be in a thermal state, it does perceive modified field correlators~\cite{Letaw1980,Kim1987}. In fact, although the field content perceived by the rotating detector remains zero, the detector still has a nonzero excitation rate~\cite{grove1983,davies1996}. This distinguishes the uniform linear and circular acceleration scenarios because in the case of uniform linear acceleration, the field content is truly changed, which constitutes the Unruh effect. However, both the Unruh effect and the response of a circularly rotating detector can be understood in terms of acceleration-induced modifications to the field correlators.  Therefore, although detecting the modified field correlators by a circularly rotating detector is not direct evidence for the standard Unruh effect, any such detection serves as a manifestation of the quantum field-theoretic non-inertial effects~\cite{bell1983,*bell1987,*unruh1998}.

Here, we have studied the GP acquired by a circularly rotating two-level atom, inside an electromagnetic cavity, interacting with the electromagnetic field in the inertial vacuum. The acceleration-induced modifications to the field correlators perceived by the atom depend on the angular frequency of the rotating atom. We have studied GP in two distinct regimes characterized by $\omega \gg \bar{\Omega}_0$ and $\omega \ll \bar{\Omega}_0$. The $\omega \ll \bar{\Omega}_0$ regime is of particular experimental interest because one of the main hindrances to the detection of acceleration-induced modifications to field correlators is that such detection requires very high accelerations.

In the $\omega \gg \bar{\Omega}_0$ regime, for $\omega \sim 10^9 \si{Hz}$ and $\Omega_0 \sim 10^7 \si{Hz}$, we have shown that the atom acquires a non-inertial GP $\sim 10^{-6} ~\si{radian}$ in $10^5$ quasi-cycles, i.e., $\sim 10^{-2}\si{s}$, while the inertial contribution remains insignificant, thereby successfully isolating the non-inertial response to the GP from the inertial one.

In general, in the $\omega \ll \bar{\Omega}_0$ regime the non-inertial GP comes out be much weaker than the inertial GP. However, we show that it is possible to make the two contributions comparable by tuning the cavity to $\bar{\Omega}_0 + \omega$ and taking a larger radius $(R)$. Specifically,  we achieve this by weakening the inertial response by tuning the cavity away from the atomic resonance.  Note that we cannot indiscriminately increase $R$ because an atom rotating on a larger radius requires a bigger cavity to encase it and a larger cavity volume suppresses the overall detector response. By allowing the atom to evolve for $\sim 10^7$ quasi-cycles, a non-inertial GP $\sim 10^{-6} ~\si{radian}$ can be acquired, which is comparable to the inertial GP acquired by the atom. This will enable the possibility of the detection of acceleration-induced modifications to field correlators with much more relaxed parameters compared to previous studies~\cite{Martinez2011,lochan2020}. 

Specifically, we have shown, in the $\omega \ll \bar{\Omega}_0$ regime, that it is possible to detect the acceleration-induced modifications to the field correlators at an acceleration $\sim 10^7 \si{m/s^2}$. Compare this with the accelerations required for the non-inertial effects to be substantial in other proposals. For example, the Unruh effect demands acceleration of the order of $10^{21} \si{m/s^2}$ if the detector transition rates are used as an observable~\cite{unruh1976}, and $10^{17}\si{m/s^2}$ if GP is used as an observable~\cite{Martinez2011}. Similarly, detecting non-inertial effects using a circularly rotating atom inside an electromagnetic cavity, by observing the atomic spontaneous decay rate, requires an acceleration of the order of $10^{12} \si{m/s^2}$~\cite{lochan2020}. Thus we demonstrate that, aided by the cavity, usage of the GP response for observing weak, but nontrivial, non-inertial effects in quantum field theory is a much sensitive and powerful tool.

\begin{acknowledgments}
	N.A. acknowledges the financial support from the University Grants Commission (UGC), Government of India, in the form of a research fellowship (Sr.~No.~2061651285). Research of K.L. is partially supported by the Startup Research Grant of SERB, Government of India (SRG/2019/002202). S.K.G. and V.M. acknowledge the financial support from the Interdisciplinary Cyber Physical Systems (ICPS) programme of the Department of Science and Technology, India (Grant No. DST/ICPS/QuST/Theme-1/2019/12). N.A. and V.M. thank G.P. Teja for useful discussions.
\end{acknowledgments}

\end{document}